\documentclass[11pt]{article}
\usepackage{epsfig}
\usepackage{mathtext}      
\usepackage{amsmath,amssymb} 
\usepackage{mathtools}
\newcommand{\be}{\begin{equation}}
\newcommand{\ee}{\end{equation}}
\newcommand{\bea}{\begin{eqnarray}}
\newcommand{\eea}{\end{eqnarray}}
\newcommand{\ba}{\begin{eqnarray*}}
\newcommand{\ea}{\end{eqnarray*}}

\def\({\left(}
\def\){\right)}

\def\f21{{_2F_1}}

\newcommand{\Fe}[2]{\Fs{#1}{#2}{z}}
\newcommand{\Foz}[2]{\Fs{#1}{#2}{\frac{1}{z}}}

\newcommand{\Fst}[2]{\Fs{#1}{#2}{-\frac{t}{s} }}

\newcommand{\Fh}[2]{\,{}_#1F_#2}
\newcommand{\Fs}[3]{\!\!\left[\begin{array}{c}#1\,;\\#2\,;\end{array}#3\right]}

\begin{document}


\title{
\vskip-2cm{\baselineskip14pt
}
\vskip2.5cm
{
Massless on-shell box integral  with arbitrary powers of propagators
}
\\}

\medskip
\vskip3.5cm

\author{
{\sc O.~V.~Tarasov}
\\
\\
{\normalsize Joint Institute for Nuclear Research,}\\
{\normalsize 141980 Dubna, Moscow Region, Russia.}}

\maketitle


\begin{abstract}

The massless one-loop box integral with arbitrary indices
in arbitrary space-time dimension $d$ is shown to 
reduce to a sum over three generalised hypergeometric functions. 
This result follows from the solution to the third order differential 
equation of  hypergeometric type. 
To derive the differential  equation, the Gr\"obner basis
technique for integrals with noninteger powers of propagators
was used.
A complete set of  recurrence relations from the Gr\"obner basis is
presented. The first several terms in the $\varepsilon =(4-d)/2$ expansion 
of the result are given.


\vskip0.5cm

\noindent
Pacs numbers: {02.30.Gp, 02.30.Ks, 12.20.Ds, 12.38.Bx} \\
Keywords: { Feynman integrals, differential equations}

\end{abstract}

\section{Introduction}

Modern  methods for evaluating  Feynman integrals can be used mainly for obtaining 
analytical results for integrals with integer powers (indices)  of propagators. 
There are no regular  algorithms for calculating integrals with noninteger powers 
of propagators. However, in some important cases one needs to evaluate integrals
with noninteger indices. For example, such integrals are needed 
when analytic regularization is exploited. Another example -  evaluation
of integrals  with massive propagators by the Mellin-Barnes technique.
Such integrals are expressible as Mellin-Barnes integrals with  integrand 
being massless integrals taken with  arbitrary indices \cite{Boos:1990rg}.
Also, compuation of multi-loop integrals with insertions of 
loop integrals with massless propagators can be reduced 
to evaluation of integrals with noninteger indices.

In this paper, we will describe a regular method for deriving 
different types of equations for integrals with noninteger 
indices. Solving these equations one can  obtain analytical
results for the required integral.
The method will be applied to
the one-loop four point massless box integral. 
This integral has been known for a long time 
with integer indices.
In our paper, we present an analytical result for this  
 integral 
with \emph{arbitrary} powers of the propagators. 

This article is organized as follows:
in Section 2, definitions and method for deriving different
types of equations for Feynman integrals with noninteger powers
of propagators are described.
In Section 3, the  differential equation and its solution for
the on-shell box integral is given.
In Section 4, different results for epsilon expansion of the result
are presented. In Section 5, the most important results of our investigation
are shortly described.

\section{Definitions and method for deriving equations}
The one-loop box type integral $I_4^{(d)}(\{\nu_j^2\},\{s_{kr}\}) $
with massless propagators to be considered in this paper is defined as 
\begin{eqnarray}
\lefteqn{I_4^{(d)}(\nu_1,\nu_2,\nu_3,\nu_4;~
s_{12},s_{23},s_{34},s_{14};~s_{24},s_{13})}
\nonumber
\\
&=&
\int \frac{d^dq}{i\pi^{d/2}}~\frac{1}{[(q-p_1)^2]^{\nu_1}
[(q-p_2)^2]^{\nu_2}[(q-p_3)^2]^{\nu_3}[(q-p_4)^2]^{\nu_4}},
\end{eqnarray}
where
\begin{equation}
\qquad  s_{ij}=p_{ij}^2,
\qquad  p_{ij}=p_i-p_j.
\end{equation}
The diagram corresponding to the integral 
is presented in Fig.~1.
\begin{figure}[h]
\begin{center}
\includegraphics[scale=0.9]{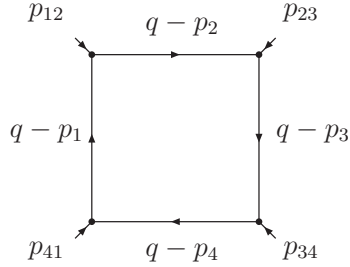}
\end{center}
\caption{\it  Diagram corresponding to the integral  
$I_4^{(d)}(\nu_1,\nu_2,\nu_3,\nu_4;~
s_{12},s_{23},s_{34},s_{14};~s_{24},s_{13}).$}
\end{figure}
In the present paper,  the on-shell version of this
integral will be considered:
\begin{equation}
I_4^{(d)}(\nu_1,\nu_2,\nu_3,\nu_4;~s_{24},s_{13})\equiv
I_4^{(d)}(\nu_1,\nu_2,\nu_3,\nu_4;~
0,0,0,0;~s_{24},s_{13}).
\end{equation}
In what follows we will use 
traditional kinematic invariants  $s$ and $t$
defined as :
\begin{equation}
s=(p_{12}+p_{41})^2,~~~~~~~~~t=(p_{12}+p_{23})^2.
\label{s_t}
\end{equation}
The analytical result for this integral when all $\nu_j=1$ 
was derived by different methods in Refs. \cite{vanNeerven:1985xr},
\cite{Bern:1993kr},
\cite{Duplancic:2000sk}, \cite{Kniehl:2009pv}. The result
for arbitrary value of $\nu_j$  was not known until now.

Parametric representations of Feynman  integrals with arbitrary 
indices are not very suitable for directly obtaining an analytical result. 
Instead, one can try to get the result
as a solution of an equation for the integral. To obtain such an
equation, we propose to use the Gr\"obner basis technique.
The idea how to derive the  Gr\"obner basis 
for integrals with noninteger powers of propagators was
outlined in Ref. \cite{Tarasov:1998nx}. Exploiting the Gr\"obner
basis technique, one can set up the differential equation for
the integral  as well as 
 recurrence relations either with respect to the
parameter of the space-time dimension $d$ or  an index $\nu_j$. 
We derived all these equations and discovered that
solving  the differential equation for 
the integral $I_4^{(d)}$ is easier than solving both types of
the recurrence relations.
For this reason we will stick to the derivation and solution
of the differential equation.

To obtain the Gr\"obner basis for $I_4^{(d)}$, we used one of the methods 
for solving the recurrence relations for
Feynman integrals proposed in \cite{Tarasov:1998nx}.
The Gr\"obner basis  consists of 12 relations. Three of these relations are
rather simple:
\begin{equation}
{\bf 3^{+}}B_3=B_1,~~~{\bf 4^+} B_3=B_2, ~~~{\bf 3^{+}}B_2={\bf 4^+}
B_1,
\end{equation}
where ${\bf j^+}$ is an operator shifting the index $\nu_j$ by one unit
\begin{equation}
{\bf j^+}I_4^{(d)}({\ldots} ,\nu_j,{\ldots} )=
I_4^{(d)}({\ldots} ,\nu_j+1,{\ldots} ).
\end{equation}
Other 9 relations from the Gr\"obner basis are given in  Appendix A.
We discovered that the
recurrence relations from this basis allow one to reduce any integral
of the type 
\begin{equation}
I_4^{(d)}(\nu_1+n_1,\nu_2+n_2,\nu_3+n_3,\nu_4+n_4,s,t)
\label{I4integershift}
\end{equation}
with integer $n_k$ to a set of three basis integrals:
\begin{eqnarray}
&&B_1  \equiv B_1(\nu_1,\nu_2,\nu_3,\nu_4,s,t)=
I_4^{(d)}(\nu_1-1,\nu_2-1,\nu_3,\nu_4-1,s,t),
\nonumber \\
&&B_2 \equiv B_2(\nu_1,\nu_2,\nu_3,\nu_4,s,t)=
I_4^{(d)}(\nu_1-1,\nu_2-1,\nu_3-1,\nu_4,s,t),
\nonumber \\
&&B_3\equiv B_3(\nu_1,\nu_2,\nu_3,\nu_4,s,t) =
I_4^{(d)}(\nu_1-1,\nu_2-1,\nu_3-1,\nu_4-1,s,t).
\label{BoxBasis}
\end{eqnarray}

Integrals of the type (\ref{I4integershift}) can be obtained by
applying product of shifting operators
\begin{equation}
({\bf 1^+})^{n_1+1}({\bf 2^+})^{n_2+1}({\bf 3^+})^{n_3+1}  ({\bf 4^+})^{n_4+1}
\end{equation}
to the basis integral
\begin{equation}
I_4^{(d)}(\nu_1-1,\nu_2-1,\nu_3-1,\nu_4-1,s,t).
\label{lowestI4}
\end{equation}
After application of a particular operator ${\bf j^+}$ to the
integral (\ref{lowestI4}) one should  use recurrence relations 
from the Gr\"obner basis. As a result  any integral of the type 
(\ref{I4integershift}) appeared in expressions
for derivatives will be reduced to a combination of three
basis integrals (\ref{BoxBasis}). The same way we can 
reduce the integral $I_4^{(d)}(\nu_1,\nu_2,\nu_3,\nu_4,s,t)$
to the basis integrals (\ref{BoxBasis}).

\section{Differential equation and its solution}
As was shown in Ref. \cite{Tarasov:1996bz}, the  derivatives of Feynman 
integrals can be written in terms of the  integrals with shifted 
dimension $d$  and changed indices of the propagators. 
To obtain an explicit formula for the derivatives of $I_4^{(d)}$,
we  will exploit its $\alpha$ parametric integral representation 
\begin{equation}
I_4^{(d)}(\nu_1,\nu_2,\nu_3,\nu_4,s,t)=
i^{\sum_{j=1}^4\nu_j-d/2-1}
\int_{0}^{\infty} {\ldots} \int_{0}^{\infty} \frac{\alpha_1^{\nu_1-1}\alpha_2^{\nu_2-1}
\alpha_3^{\nu_3-1}\alpha_4^{\nu_4-1}}
{\Gamma(\nu_1)\Gamma(\nu_2)\Gamma(\nu_3)\Gamma(\nu_4)}
\frac{\{d\alpha\}}{D^{d/2}}
e^{iQ/D-i\epsilon},
\end{equation}
where $\{d\alpha\}=d\alpha_1{\ldots} d\alpha_4$ and
\begin{eqnarray}
&&Q=\alpha_1\alpha_3t + \alpha_2\alpha_4 s,
\nonumber  \\
&&D=\alpha_1+\alpha_2+\alpha_3+\alpha_4.
\label{QiDpolynoms}
\end{eqnarray}
According to the method of  Ref.\cite{Tarasov:1996bz},
explicit expressions for  the derivatives of the integral $I_4^{(d)}$ are
determined by the  polynomial $Q$ in Eq. (\ref{QiDpolynoms}) and they read:
\begin{eqnarray}
\label{I4derivative_t}
&&\frac{\partial}{\partial t}I_4^{(d)}(\nu_1,\nu_2,\nu_3,\nu_4,s,t)
=\nu_1\nu_3 I_4^{(d+2)}(\nu_1+1,\nu_2,\nu_3+1,\nu_4,s,t),
\\
&&\frac{\partial}{\partial s}I_4^{(d)}(\nu_1,\nu_2,\nu_3,\nu_4,s,t)
=\nu_2\nu_4 I_4^{(d+2)}(\nu_1,\nu_2+1,\nu_3,\nu_4+1,s,t).
\label{I4derivative_s}
\end{eqnarray}
Exploiting equations (\ref{I4derivative_t}), (\ref{I4derivative_s}), one
can also express higher derivatives of the integral $I_4^{(d)}$ in terms
of the integrals with shifted dimension and indices.

In order to obtain the differential equation, we expressed the 
basis integrals in terms of the integral itself and its first and 
second derivatives. Substituting these expressions 
into the result for the third derivative leads to the
following differential equation for the integral
$I_4^{(d)}(\nu_1,\nu_2,\nu_3,\nu_4,s,t) \equiv I^{(d)}_4(t)  $:
\begin{eqnarray}
&&4t^2(s+t)~\theta^3I^{(d)}_4(t)
\nonumber \\
&&
-2t[2(d-2 \nu_1-\nu_2-2 \nu_3 - \nu_4-3)s+(d-6-4 \nu_1-2 \nu_2-4 \nu_3-2 \nu_4) t]
~\theta^2 I^{(d)}_4(t)
\nonumber \\
&&+[2(2 \nu_1 \nu_2+2 \nu_1 \nu_4+2 \nu_2+2 \nu_4+2 \nu_3 \nu_4
+2+4 \nu_3- d \nu_1- \nu_3 d
\nonumber \\
&&
+2 \nu_2 \nu_3+6 \nu_1 \nu_3+2 \nu_3^2- d+2 \nu_1^2+4 \nu_1) t
\nonumber \\
&&+s (d-2\nu+2\nu_2-2) (d-2\nu+2\nu_4 -2)]~\theta I^{(d)}_4(t)
 -2 \nu_1 \nu_3 (d-2\nu)~I^{(d)}_4(t)=0,
\label{dif_equ_I4}
\end{eqnarray}
where
\begin{equation}
\theta = \frac{d}{dt}.
\end{equation}
and
\begin{equation}
\nu=\nu_1+\nu_2+\nu_3+\nu_4.
\end{equation}
Similar prescription can be used to obtain recurrence relations
with respect to an index $\nu_j$ or $d$.

Solution of the differential equation (\ref{dif_equ_I4}) which is valid at 
small $t$ reads:
\begin{eqnarray}
&&I^{(d)}_4(t)=\frac{ C_1(\nu_1,\nu_2,\nu_3,\nu_4)}{ s^{\nu-\frac{d}{2} }}
 \Fh32\Fst{\nu_1, \nu_3, \nu-\frac{d}{2}}{ \nu-\nu_4+1-\frac{d}{2}, 
 \nu-\nu_2+1-\frac{d}{2}}
\nonumber \\
&&~~~~~~
+ \frac{C_2(\nu_1,\nu_2,\nu_3,\nu_4)}{t^{\nu-\frac{d}{2}-\nu_4}~
  s^{\nu_4}}
\Fh32\Fst{\nu_4, \frac{d}{2}-\nu_1-\nu_2, \frac{d}{2}-\nu_3-\nu_2}
	 {1+\nu_4-\nu_2, \frac{d}{2}-\nu+\nu_4+1}
\nonumber \\
&&~~~~~~
     +\frac{C_3(\nu_1,\nu_2,\nu_3,\nu_4)}
     {t^{\nu-\frac{d}{2}-\nu_2}s^{\nu_2}}
 \Fh32\Fst{\nu_2, \frac{d}{2}-\nu_4-\nu_1, \frac{d}{2}-\nu_4-\nu_3}
	 {1-\nu_4+\nu_2, \frac{d}{2}-\nu+ \nu_2+1},
\label{Solution}
\end{eqnarray}
where
\begin{equation}
 \Fh32\Fe{a_1, a_2, a_3}{b_1,b_2}=\sum_{k=0}^{\infty}
 \frac{(a_1)_k (a_2)_k (a_3)_k}{(b_1)_k (b_2)_k} \frac{z^k}{k!},
\end{equation}
and $(a_j)_k=\Gamma(a_j+k)/\Gamma(a_j)$.
Arbitrary constants $C_j$ can be determined from boundary conditions
and symmetric relations of the integral with respect to indices.
The coefficient $C_1(\nu_1,\nu_2,\nu_3,\nu_4)$ 
can be found from the value of the integral at $t=0$:
\begin{equation}
I_4^{(d)}(t=0) = i^d s^{d/2-\nu}
\frac{\Gamma\left(\nu-\frac{d}{2}\right)
\Gamma\left(\frac{d}{2}-\nu+\nu_2 \right)
\Gamma\left(\frac{d}{2}-\nu + \nu_4\right)}
{\Gamma(\nu_2)\Gamma(\nu_4) \Gamma(d-\nu)},
\label{boundary_t_zero}
\end{equation}
Setting $t=0$ in Eq.(\ref{Solution}) leads to the relation 
\begin{equation}
C_1(\nu_1,\nu_2,\nu_3,\nu_4)=s^{\nu-d/2}I_4^{(d)}(t=0)
=i^d
\frac{\Gamma\left(\nu-\frac{d}{2}\right)
\Gamma\left(\frac{d}{2}-\nu+\nu_2 \right)
\Gamma\left(\frac{d}{2}-\nu + \nu_4\right)}
{\Gamma(\nu_2)\Gamma(\nu_4) \Gamma(d-\nu)}.
\end{equation}

In order to find $C_2$ and $C_3$, we first transformed the result (\ref{Solution})
into the region valid for large $t$ by using the formula for
analytic continuation of $_3F_2$ functions:
\begin{eqnarray}
&&\Fh32\Fe{a_1,a_2,a_3}{b_1,b_2}=
\frac{\Gamma(b_1) \Gamma(b_2)}{\Gamma(a_1) \Gamma(a_2) \Gamma(a_3)}
\nonumber \\
&&\times
\left\{\frac{
\Gamma(a_2-a_1)\Gamma(a_3-a_1) \Gamma(a_1)}
{\Gamma(b_1-a_1)\Gamma(b_2-a_1) (-z)^{a_1}}
 \Fh32\Foz{a_1,1-b_1+a_1,1-b_2+a_1}{1-a_2+a_1,1-a_3+a_1}
\right.
\nonumber \\
&&
+\frac{\Gamma(a_1-a_2)\Gamma(a_3-a_2)\Gamma(a_2)}
{\Gamma(b_1-a_2)\Gamma(b_2-a_2)~(-z)^{a_2}}
 \Fh32\Foz{a_2,1-b_1+a_2,1-b_2+a_2}{1-a_1+a_2,1-a_3+a_2}
\nonumber \\
&&\left.
+\frac{
\Gamma(a_1-a_3) \Gamma(a_2-a_3)\Gamma(a_3)}{\Gamma(b_1-a_3)\Gamma(b_2-a_3)~(-z)^{a_3}}
 \Fh32\Foz{a_3,1-b_1+a_3,1-b_2+a_3}{1-a_1+a_3,1-a_2+a_3} \right\}.
\label{F32_analytic}
\end{eqnarray}
Taking into account  symmetries of the integral with respect to $\nu_j$:
\begin{eqnarray}
&&
I_4^{(d)}(\nu_1,\nu_2,\nu_3,\nu_4,s,t) =
I_4^{(d)}(\nu_3,\nu_2,\nu_1,\nu_4,s,t)=
I_4^{(d)}(\nu_1,\nu_4,\nu_3,\nu_2,s,t),
\nonumber \\
&&
I_4^{(d)}(\nu_1,\nu_2,\nu_3,\nu_4,s,t)=
I_4^{(d)}(\nu_3,\nu_4,\nu_1,\nu_2,t,s),
\label{symmetries}
\end{eqnarray}
and the value of the integral at $s=0$
\begin{equation}
I_4^{(d)}(s=0) = i^d t^{d/2-\nu}
\frac{\Gamma\left(\nu-\frac{d}{2}\right)
\Gamma\left(\frac{d}{2}-\nu+\nu_1 \right)
\Gamma\left(\frac{d}{2}-\nu + \nu_3\right)}
{\Gamma(\nu_1)\Gamma(\nu_3) \Gamma(d-\nu)},
\label{boundary_s_zero}
\end{equation}
the following formulae for  $C_2$ and  $C_3$  were derived:
\begin{eqnarray}
&&C_2(\nu_1,\nu_2,\nu_3,\nu_4)=i^d
 \frac{\Gamma\left(\nu-\nu_4 -\frac{d}{2} \right)
 \Gamma\left(\frac{d}{2}-\nu_3-\nu_2\right) \Gamma\left(\nu_2-\nu_4\right)
  \Gamma\left(\frac{d}{2}-\nu_1-\nu_2\right)}
 {\Gamma\left(\nu_1\right) \Gamma\left(\nu_3\right) \Gamma\left(\nu_2\right)
 \Gamma\left(d-\nu\right)},
\nonumber \\
&& \nonumber \\
&&C_3(\nu_1,\nu_2,\nu_3,\nu_4)=C_2(\nu_1,\nu_4,\nu_3,\nu_2).
\end{eqnarray}

Symmetric relations  (\ref{symmetries}) can be easily seen from the $\alpha$ parametric
representation of $I_4^{(d)}$.
From Eqs.(\ref{symmetries}) and hypergeometric representation (\ref{Solution})
also the following  relations follow:
\begin{eqnarray}
&&C_2(\nu_1,\nu_2,\nu_3,\nu_4)=C_2(\nu_3,\nu_2,\nu_1,\nu_4),
\nonumber \\
&&C_2(\nu_1,\nu_2,\nu_3,\nu_4)=C_3(\nu_1,\nu_4,\nu_3,\nu_2).
\nonumber \\
&&C_3(\nu_1,\nu_2,\nu_3,\nu_4)=C_2(\nu_1,\nu_4,\nu_3,\nu_2).
\end{eqnarray}

\section{Epsilon expansion of $I_4^{(d)}$}
In this section we present several terms in the  $\varepsilon$ 
expansion of the on-shell integral $I_4^{(d)}$ for the case when
\begin{equation}
\nu_1=1+a_1\varepsilon,~~\nu_2=1+a_2\varepsilon,~~
\nu_3=1+a_3\varepsilon,~~\nu_4=1+a_4\varepsilon.~~
\label{nu_definition}
\end{equation}
Epsilon expansion of hypergeometric functions was described in Refs.
\cite{Huber:2005yg},\cite{Huber:2007dx},\cite{Huber:2008mz},
\cite{Kalmykov:2007dk},\cite{Kalmykov:2007pf}, \cite{Huang:2012qz}.
The result up to $O(\varepsilon^2)$ terms reads:
\begin{eqnarray}
&&- \frac{z}{a} (1+a_1+a_4)(1+a_3+a_4)(1+a_2+a_3)(1+a_1+a_2)
s^{(\nu-d/2)} I_4^{(d)}(t)
\nonumber\\
&&
=
\frac{a}{\varepsilon^2}
 -\Big[(1+a_3+a_4) (1+a_2+a_3)+ a~ a_1\Big] \frac{\ln(-z)}{\varepsilon}
\nonumber \\
&&+\frac12 \Big[(a_1+a_3) (1+a_3+a_4) (1+a_2+a_3)+a~ a_1^2\Big]
\ln^2(-z)
\nonumber \\
&&
-\frac{\pi^2}{6}\Big[(a_2+1) (1+a_4) (2 a_2+3+2 a_4)+(3 a_2+7+3 a_4) a_1 a_3
\nonumber \\
&&~~
+(3 a_2 a_4+a_2^2+5 a_4+a_4^2+5+5 a_2) (a_1+a_3)
\nonumber \\
&&~~
+(2 a_1+a_2+2+a_4) a_3^2
  +(2 a_3+a_2+2+a_4) a_1^2\Big]
\nonumber \\
&&+\varepsilon\Big\{
(1+a_1+a_4) (1+a_3+a_4) (1+a_2+a_3) (1+a_1+a_2)
 \Big[
 -1/2 \ln(1-z) \ln(-z)^2
\nonumber \\
&&
 - \ln(-z) {\rm Li}_2(z)
 -\frac{\pi^2}{2} \ln(1-z) + {\rm Li}_3(z)\Big]
+\frac16 S_1 \ln(-z)^3 +\frac{\pi^2}{6} S_2  \ln(-z) +S_3 \zeta_3
\Big\}
\nonumber \\
&&~~~~~~~~~~~~~~~~~~~~~~~~~~~~~~~~~~~~~~~~~~~ + O(\varepsilon^2),
\label{epsilon_expansion}
\end{eqnarray}
where $S_1$,$S_2$, $S_3$ are given in Appendix B and
\begin{equation}
a=2 + a_1 + a_2 + a_3 + a_4.
\end{equation}
When all $a_k=0$ for the first three terms in the expansion, we found complete 
agreement with the result given in Ref.\cite{Bern:1993kr}.

Just for completeness we present here the formula for $I_4^{(d)}$
which can be used in the frame of analytical regularization.
In this case it is assumed that $d=4$ and indices are defined as
in (\ref{nu_definition}).  The result of
$\varepsilon$ expansion including the constant term reads:
\begin{eqnarray}
&&
- \frac{z}{b} (a_1+a_4)(a_3+a_4)(a_2+a_3)(a_1+a_2)
s^{(\nu-d/2)} I_4^{(d)}(t)
\nonumber\\
&&
=
\frac{b}{\varepsilon^2}
 -\Big[(a_3+a_4) (a_2+a_3)+ b~ a_1\Big] \frac{\ln(-z)}{\varepsilon}
\nonumber \\
&&
-\frac{\pi^2}{6}\Big[(a_2+a_4)(2a_2 a_4+a_1^2+a_3^2+3a_1a_3)
+(a_1+a_3)(a_2^2+3a_2a_4+a_4^2+2a_1a_3)
\Big]
\nonumber \\
&& ~~+\frac12
 \Big[(a_1+a_3) (a_3+a_4) (a_2+a_3)+b~ a_1^2\Big]\ln(-z)^2
 +O(\varepsilon),
\end{eqnarray}
where
\begin{equation}
b=a-2.
\end{equation}

\section{Conclusions}
In conclusion, we shortly summarize the most important results of the paper.
First of all, we outlined a regular algorithm for deriving
analytical results for Feynman integrals with arbitrary noninteger
powers of propagators.
We illustrated that the algorithm perfectly works for
a rather complicated integral. It can be used  not only for deriving
differential equations but also for deriving  recurrence relations
with respect to space-time dimension $d$ or an index $\nu_j$ 
for integrals. From these equations one can choose
an equation the easiest for obtaining an analytical result.
The method can be applied to massless integrals as well as 
integrals with massive propagators.

Detailed description of our method and its apllication to
other integrals will be presented in our future publications.

\vspace{1cm}
{\it Acknowledgments}
This work was supported by the German Science Foundation (DFG)
within the Collaborative Research Center 676 "Particles, Strings and the Early Universe". 

\section{Appendices}
\vspace{-2mm}
{\bf Appendix A}\\
Recurrence relations from the Gr\"obner basis for the integral
$I_4^{(d)}(\nu_1,\nu_2,\nu_3,\nu_4,s,t)$.
\begin{eqnarray}
&&{\bf 1^+} B_1=
I_4^{(d)}(\nu_1,\nu_2-1,\nu_3,\nu_4-1,s,t)\nonumber \\
&&= -\frac{(d+2-\nu)(d-2\nu+4+2\nu_4)}
{ t(\nu_1-1) (2-2\nu_1-2\nu_2+d)} B_1
+\frac{2(\nu_4-1) (d+2-\nu)}{ t (\nu_1-1) (2-2\nu_1-2\nu_2+d)} B_2,
\\
\nonumber \\
&&{\bf 2^+} B_1=
I_4^{(d)}(\nu_1-1,\nu_2,\nu_3,\nu_4-1,s,t)
\nonumber \\
&&
= -\frac{(t(d+2\nu_1-2\nu+4)(2-2\nu_1-2\nu_2+d)
 +2(\nu_2-\nu_4)(d+2\nu_4 -2\nu+4)s)(d+2-\nu)}
  {(d-2\nu_3-2\nu_2)(\nu_2-1) t s (2-2\nu_1-2\nu_2+d)}B_1
  \nonumber \\
  &&
   -\frac{ (d-2\nu_1-2\nu_4+2)(d-2\nu+4+2\nu_4)(\nu_4-1)(d+2-\nu)}
   {(\nu_2-1)t (d-2\nu_3-2\nu_2) (\nu_3-1) (2-2\nu_1-2\nu_2+d)}B_2
   \nonumber \\
   &&
   -\frac{(d-2 \nu+8) (d+3-\nu)(d+2-\nu)}
   {(\nu_2-1)(\nu_3-1)(d-2\nu_3-2\nu_2) s t}B_3,
\end{eqnarray}
   \begin{eqnarray}
   &&{\bf 3^+}B_1=
   I_4^{(d)}(\nu_1-1,\nu_2-1,\nu_3+1,\nu_4-1,s,t) 
   \nonumber \\
   &&
   =
   -\frac{((d-2\nu_4-2\nu_3)(d+2\nu_4-2\nu+4) s
    -2t(\nu_1-\nu_3-1)(d+2\nu_1-2\nu+4)) (d+2-\nu)}
    {(d-2 \nu_4-2\nu_3)(d-2\nu_3-2\nu_2) s t \nu_3}B_1
    \nonumber \\
    &&
    +\frac{2(\nu_1-2)(d-2\nu_1-2\nu_4+2)(\nu_4-1)(d-\nu +2)}
    { t (d-2\nu_4-2\nu_3) \nu_3 (d-2\nu_3-2\nu_2) (\nu_3-1)} B_2
    \nonumber \\
    &&
     +\frac{2 (\nu_1-\nu_3-1)(d-2\nu+8)(d-\nu+3)(d-\nu +2)}
      {(\nu_3-1) (d-2\nu_3-2\nu_2) \nu_3 (d-2\nu_4-2\nu_3) s t}B_3,
\\
      \nonumber \\
      &&{\bf 4^+} B_1=
      I_4^{(d)}(\nu_1-1,\nu_2-1,\nu_3,\nu_4,s,t)
      \nonumber \\
      &&
      = 
       - \frac{(d-2\nu+4+2\nu_1)(d+2-\nu)}{ (\nu_4-1) s
       (d-2\nu_4-2\nu_3)}B_1
       -\frac{(4+d+2\nu_2-2\nu )(d+2-\nu)}{t(d-2\nu_4-2\nu_3)
       (\nu_3-1)}B_2
       \nonumber \\
       &&
        -\frac{(d-2\nu+8)(d+3-\nu)(d+2-\nu)}
	 {(d-2\nu_4-2\nu_3) s t (\nu_4-1)(\nu_3-1)}B_3,
    \end{eqnarray}
  \begin{eqnarray}
	 &&{\bf 1^+} B_2=
	 I_4^{(d)}(\nu_1,\nu_2-1,\nu_3-1,\nu_4,s,t) 
	 \nonumber \\
	 &&= 
	  -\frac{ (d-2 \nu+2 \nu_3+4) (\nu_3-1) (d+2-\nu) (d+2-2
	  \nu_3-2 \nu_2)}
	   {(\nu_4-1)(d-2 \nu_1-2 \nu_4)(2-2 \nu_1-2
	   \nu_2+d)s(\nu_1-1)} B_1
	   \nonumber \\
	   &&
	    - \frac{(2 t (\nu_1-\nu_3) (d+2\nu_3-2 \nu+4)
	     + (4+2 \nu_2-2 \nu+d) (2-2\nu_1-2\nu_2+d) s)(d+2-\nu)}
	     {(2-2 \nu_1-2 \nu_2+d) t s (d-2 \nu_1-2\nu_4)(\nu_1-1)} B_2
	     \nonumber \\
	     &&
	     -\frac{ (d-2 \nu+8) (d+3-\nu) (d+2-\nu)}
	     {s t (\nu_1-1) (\nu_4-1) (d-2 \nu_1-2 \nu_4)} B_3,
\\
\nonumber \\
	     &&{\bf 2^+}B_2=
	     I_4^{(d)}(\nu_1-1,\nu_2,\nu_3-1,\nu_4,s,t) 
	     \nonumber \\
	     &&
	     =
	      \frac{2 (d+2-\nu)(\nu_3-1)}{s (\nu_2-1)
	      (2-2\nu_1-2\nu_2+d)} B_1
	      -\frac{(d+2-\nu) (d+2\nu_3 -2\nu+4)}
	      { s(\nu_2-1)(2-2\nu_1-2\nu_2+d)}B_2,
	      \end{eqnarray}
       	 \begin{eqnarray}
	      &&{\bf 4^+}B_2=
	      I_4(\nu_1-1,\nu_2-1,\nu_3-1,\nu_4+1,s,t)
	      \nonumber \\
	      &&
	      = 
	      \frac{2(\nu_3-1)(\nu_2-2)(d+2-2\nu_3-2\nu_2)(d-\nu+2)}
	      {(d-2\nu_1-2\nu_4) \nu_4(d-2\nu_4-2\nu_3) s (\nu_4-1)}B_1
	      \nonumber \\
	      &&
	      -\frac{(d-\nu+2)(t(d-2\nu_4-2\nu_3)(d+2\nu_3 -2\nu+4)
	       -2 (\nu_2-\nu_4-1)(4+2\nu_2-2\nu+d)s)}
	       {(d-2\nu_1-2\nu_4) (d-2\nu_4-2\nu_3) s \nu_4 t}B_2
	       \nonumber \\
	       &&
	       + \frac{2 (\nu_2-\nu_4-1) (d-2\nu+8) (d-\nu +3)(d-\nu+2)}
	       {(d-2\nu_4-2\nu_3) s t (\nu_4-1) (d-2\nu_1-2\nu_4)\nu_4}B_3,
             \\
	       &&{\bf 1^+} B_3 = I_4^{(d)}(\nu_1,\nu_2-1,\nu_3-1,\nu_4-1,s,t)
	       \nonumber \\
	       &&
	       = \frac{(d+2-2\nu_3-2\nu_2) (\nu_3-1)}{(\nu_1-1) (2-2
	       \nu_1-2 \nu_2+d)} B_1
	       +\frac{2(\nu_1-\nu_3)(\nu_4-1)}{(\nu_1-1)(2-2\nu_1-2
	       \nu_2+d)}B_2,
	       \\
	       &&{\bf 2^+}B_3=
	       I_4^{(d)}(\nu_1-1,\nu_2,\nu_3-1,\nu_4-1,s,t) 
	       \nonumber \\
	       &&= \frac{2(\nu_3-1)(\nu_2-\nu_4)}
	       {(\nu_2-1) (2-2\nu_1-2\nu_2+d)}B_1
	       +\frac{(\nu_4-1) (d-2\nu_1-2\nu_4+2) }
	       {(\nu_2-1)(2-2\nu_1-2\nu_2+d)} B_2,
	       \end{eqnarray}
	       
	       \begin{eqnarray}
	       &&
	       I_4^{(d+2)}(\nu_1-1,\nu_2-1,\nu_3-1,\nu_4-1,s,t)
	       \nonumber \\
	       &&
	        = \frac{( 2\nu_2 t-4t+(d+2\nu_2-2\nu+6)s)(\nu_3-1) s t
		(d+2-2\nu_3-2\nu_2)}
		 { 2 (d+3-\nu)(d-2\nu+10)(d+5-\nu)(d-\nu+4)(t+s)}B_1
		 \nonumber \\
		 &&
		 +\frac{(d-2\nu_1-2\nu_4+2)( t (d+2\nu_1-2\nu+6)
		  +2\nu_1 s-4 s) t s (\nu_4-1)}
		    {2 (d+3-\nu) (d+10-2\nu)(d+5-\nu)(d-\nu+4)(t+s)}B_2
		    \nonumber \\
		    &&~~~~~
		    +\frac{1} {2(d-\nu+4)(d+10-2\nu)(d-\nu+5)(t+s)}
		    \nonumber \\
		    &&\times[
		     t^2(d+2\nu_1-2\nu+6)(d+2\nu_3-2\nu+6)+t(d^2-2d\nu
		     +6d 
		      + 4\nu_1\nu_3-8+4\nu_2\nu_4) s
		      \nonumber \\
		      &&~~~~~~ 
		       +(d+2\nu_2-2\nu+6)
		        (d+2\nu_4 -2\nu+6)s^2]
			B_3.
			\end{eqnarray}

{\bf Appendix B}\\

Some coefficients in the $\varepsilon$ expansion
of $I_4^{(d)}$ given in Eq. (\ref{epsilon_expansion}).
\begin{eqnarray}
&&S_1=
1 + 2(a_1+a_2+a_3+a_4)
+a_2^2+a_4^2 + 4 a_2 a_4+3 a_3 (a_2+a_4)
+3(a_3+a_2+a_4) a_1
\nonumber \\
&&
- a_3 (2 a_3^2-a_4^2)+ a_4 (a_4+3 a_3) a_1
+2 a_4 (2 a_3+a_4) a_2+ (a_1+2 a_4+a_3) a_2^2
\nonumber \\
&&
+(4a_4+3 a_3) a_1 a_2-2 a_1^3
+(a_3+a_4) (a_2+a_3) ((a_2 a_4-a_3^2)
\nonumber \\
&&+(a_2-a_3+a_4) a_1)
-(a_2+a_3+a_4) a_1^3 - a_1^4,
\end{eqnarray}

\begin{eqnarray}
&&S_2=
3 + 8 (a_1+a_3)+6(a_2+a_4)
+7 (a_1^2+a_3^2)+(17 a_3+12 a_2+12 a_4) a_1
\nonumber \\
&&+3a_2^2+12 (a_4+a_3) a_2+3 a_4^2+12 a_4 a_3
+(6a_4+4 a_3+4 a_1) a_2^2
\nonumber \\
&&
+(6a_4^2+16 (a_1+a_3) a_4+16 a_1 a_3+7 a_3^2
+7 a_1^2) a_2
\nonumber \\
&&
+(16 a_1 a_3+7 a_3^2+ 7 a_1^2) a_4
+ (a_1+a_3) (2 a_1^2+9 a_1 a_3+2 a_3^2+4 a_4^2 )
\nonumber \\
&&
+ (a_3+a_4) (a_2+a_3) (a_2 a_3+3 a_2 a_4+a_3 a_4)
+(2 a_3^3+9 a_2 a_3 a_4+3 a_3 a_4^2+
3 a_3 a_2^2
\nonumber \\
&&
+4 a_2 a_4^2+4 a_2^2 a_4
+5 a_4 a_3^2+5 a_2 a_3^2) a_1
+(a_2+2 a_3+a_4) a_1^3
\nonumber \\
&&
+ (a_2^2+a_4^2+4 a_3^2+5 a_3 a_2+5 a_2 a_4+5 a_4 a_3) a_1^2,
\end{eqnarray}

\begin{eqnarray}
&&S_3=
-5-11 (a_1+a_2+a_3+a_4)
-8 (a_1^2+a_2^2+a_3^2+a_4^2)
 -17 (a_1+a_3) (a_2+a_4)
\nonumber\\
&&
-19 (a_2 a_4+a_1 a_3)
-2 (a_1^3+a_2^3+a_3^3+a_4^3)+(-8 a_2-10 a_3-8 a_4) a_1^2
\nonumber \\
&&
+(-8 a_2^2-19 (a_3+a_4) a_2-10 a_3^2-19 a_4 a_3-8 a_4^2) a_1
-(8 a_3+10 a_4) a_2^2
\nonumber \\
&&
- (8 a_3^2+19 a_3 a_4+10 a_4^2) a_2-8 a_3 a_4 (a_3+a_4)
- a_3a_4(a_3+a_4)^2
\nonumber \\
&&
-(a_2+2 a_3+a_4)a_1 ( a_1^2+(a_3+2 a_4+2 a_2) a_1
+a_2^2+2 a_2 a_3+4 a_2 a_4+a_3^2+2 a_3 a_4+a_4^2)
\nonumber \\
&&-a_2a_3(a_2+a_3)^2
-(a_2+a_4) (2 a_2+5 a_3) a_2 a_4
-4 a_2 a_4 a_3^2-2 a_2 a_4^3
.
\end{eqnarray}

\end{document}